\documentclass[prd]{revtex4}
\usepackage{graphicx}

\begin{document}

\title{Perturbative -- nonperturbative connection in
quantum mechanics and field theory}

\author{Gerald V. Dunne}

\affiliation{Department of Physics, University of Connecticut,  Storrs
CT 06269, USA}

\begin{abstract}
On the occasion of this ArkadyFest, celebrating Arkady Vainshtein's
$60^{\rm th}$ birthday, I review some selected aspects of the connection
between perturbative and nonperturbative physics, a subject to which
Arkady has made many important contributions. I first review this
connection in quantum mechanics, which was the subject of Arkady's very
first paper. Then I discuss this issue in relation to effective actions
in field theory, which also touches on Arkady's work on operator product
expansions. Finally, I conclude with a discussion of a special quantum
mechanical system, a quasi-exactly solvable model with energy-reflection
duality, which exhibits an explicit duality between the perturbative and
nonperturbative sectors, without invoking supersymmetry.
\end{abstract}

\maketitle

\section{Divergence of perturbation theory}

\centerline{\fbox{\shortstack{\\``The majority of nontrivial theories 
are seemingly unstable at some\\
phase of the coupling constant, which leads to the asymptotic\\ nature of
the perturbative series.''\\~
A. Vainshtein, 1964 \cite{arkady}}}}
\medskip

In this talk I review some aspects of the historical development of the
connection between perturbative and nonperturbative physics. It is
particularly appropriate to look back on this subject on the occasion of
Arkady Vainshtein's $60^{\rm th}$ birthday, because this has been a central
theme of many of Arkady's great contributions to theoretical physics. In
fact, in his very first physics paper \cite{arkady}, now almost 40 years
ago, Arkady made a fundamental contribution to this subject. This paper
was published as a Novosibirsk report and so has not been widely
circulated, especially in the West. For this ArkadyFest, Misha Shifman
has made an English translation of this paper, and both the original
Russian and the translation are reprinted in these Proceedings.

The physical realization of the possibility of the divergence of
perturbation theory is usually traced back to a profound and influential
paper by Dyson \cite{dyson}, in which he argued that QED perturbation theory
should be divergent. Dyson's argument goes like this: a physical quantity
in QED, computed using the standard rules of renormalized QED perturbation
theory, is expressed as a perturbative series in powers of the fine
structure constant, $\alpha=\frac{e^2}{4\pi}$:
\begin{eqnarray}
F(e^2)=c_0+c_2 e^2+c_4 e^4+\dots
\label{series}
\end{eqnarray}
Now, suppose that this perturbative expression is convergent. This means
that in some small disc-like neighborhood of the origin, $F(e^2)$ has a
well-defined convergent approximation. In particular, this means that
within this region, $F(-e^2)$ also has a well-defined convergent
expansion. Dyson then argued on physical grounds that this cannot be the
case, because if $e^2<0$ the vacuum will be unstable. This, he argued, is
because with $e^2<0$ like charges attract and it will be energetically
favorable for the vacuum to produce $e^+ e^-$ pairs which coalesce into
like-charge blobs, a runaway process that leads to an unstable state:
\medskip

\centerline{\fbox{\shortstack{\\``Thus every physical state is
unstable against the spontaneous creation\\ of large numbers of
particles.
Further, a system once in a pathological\\ state will not remain
steady;  there will be a rapid creation of more\\ and more
particles, an explosive disintegration of the vacuum\\ by
spontaneous polarization.''\\~ F. J. Dyson, 1952
\cite{dyson}}}}
\medskip
\noindent
The standard QED
perturbation theory formalism breaks down in such an unstable vacuum, which
Dyson argued means that $F(-e^2)$ cannot be well-defined, and so the
original perturbative expansion (\ref{series}) cannot have been convergent.

Dyson's argument captures beautifully an essential piece of physics, namely
the deep connection between instability and the divergence of perturbation
theory. The argument is not mathematically rigorous, and does not
{\it prove} one way or another whether QED perturbation theory is
convergent or divergent, or analytic or nonanalytic. However, it is
nevertheless very suggestive, and has motivated many subsequent studies
in both quantum mechanics and quantum field theory.

At roughly the same time, C. A. Hurst \cite{hurst} and W. Thirring
\cite{thirring} (see also A. Petermann \cite{petermann}) showed by
explicit computation that perturbation theory diverges in  scalar
$\phi^3$ theory. Both Hurst and Thirring found lower bounds on the
contribution of Feynman graphs at a given order of perturbation theory,
and showed that these lower bounds were themselves factorially divergent.
Hurst used the parametric representation of an irreducible, renormalized
and finite
$\phi^3$ Feynman graph, to show that the magnitude of this graph was
bounded below:
\begin{eqnarray}
|I|\geq \frac{n^{-n+3/2-E/2}\, e^{2n-2}\, \pi^{-(n-E+5)/2}}{(2\pi)^F\,
2^{n+1/2-3E/2}\, 3^{4n-7/2} \, \gamma^{(n+E-4)/2}}\,\, \lambda^n
\label{hurstbound}
\end{eqnarray}
Here $n$ is the loop order, $\lambda$ is the cubic coupling constant, $E$ is
the number of external lines, $F=\frac{1}{2}(3n-E)$ is the number of
internal lines, and $\gamma$ is a constant depending on the external
momenta. This lower bound is found by  clever rearrangements of the
parametric representation, together with the identity
\begin{eqnarray}
\prod_{i=1}^F\left(\frac{1}{p_i^2+\kappa^2}\right) \geq
\frac{F^F}{\left(\sum_{i=1}^F p_i^2 +F\, \kappa^2\right)^F}
\end{eqnarray}
The second important piece of the argument is to show that there are no
sign cancellations which would prevent this lower bound from a typical
graph from being used to obtain a lower bound on the total contribution at
a given order. This requires some  technical caveats -- for example,
for a two-point function one requires $p^2 < m^2$. The final piece of
Hurst's argument is the fact that the number of distinct Feynman diagrams
at $n^{\rm th}$ loop order grows like $(\frac{n}{2})!\, n!$. 

Together, the lower bound (\ref{hurstbound}), the nonalternation of
the sign, and the rapid growth of the number of graphs, lead to a
lower bound for the total contribution at n-loop order (with E external
lines):
\begin{eqnarray}
\sum I \geq C^n \, n^{n/2+5/2-E/2}\, \lambda^n
\label{hurstfinal}
\end{eqnarray}
Here $C$ is a finite constant, independent of $n$. Therefore, Hurst
concluded that in $\lambda \phi^3$ theory, perturbation theory diverges
for \underline{any} coupling $\lambda$. He also suggests that a similar
argument should hold for $\lambda \phi^4$ theory, and comments:
\vskip 1mm

\centerline{\fbox{\shortstack{``If it be granted that the perturbation
expansion does not lead to a\\ convergent series in the coupling
constant for all theories which can\\ be renormalized, at least, then a
reconciliation is needed between this\\ and the
excellent agreement found in electrodynamics between\\ experimental
results and low-order calculations. It is suggested that this\\
agreement is due to the fact that the S-matrix expansion is to be\\
interpreted as an asymptotic expansion in the fine-structure constant
...''\\~ C. A. Hurst, 1952 \cite{hurst}}}}
\medskip

Thirring's argument \cite{thirring} was similar in spirit, although he
concentrated on the $\phi^3$ self-energy diagram. Thirring found a set of
graphs that were simple enough that their contribution could be estimated
and bounded below, while plentiful enough that they made a divergent
contribution to the perturbative series. He noted that the proof relied
essentially on the fact that certain terms always had the same sign, and
traced this fact to the hermiticity of the interaction. He found the
following (weaker) lower bound, valid for $p^2<m^2$ :
\begin{eqnarray}
\Delta(p^2)\geq \sum_n C(p^2)\, \left(\frac{\lambda\, e}{4\pi
m 3^{5/2}}\right)^n \, \frac{(\frac{n}{2}-2)!}{n^2}
\label{thirringbound}
\end{eqnarray}
Thirring concluded that there was no convergence for any $\lambda$. His
final conclusion was rather pessimistic:
\smallskip

\centerline{\fbox{\shortstack{\\``To sum up, one can say that the
chances for quantized fields to\\ become a mathematically consistent
theory are rather slender.''\\~ W. Thirring, 1953 \cite{thirring}}}}
\medskip

\noindent
These results of Dyson, Hurst and Thirring, provide the
backdrop for Arkady's first paper \cite{arkady}, ``Decaying systems and
divergence of perturbation theory'', written as a young student beginning
his PhD at Novosibirsk. I encourage the reader to read Arkady's paper --
it is simple but deep. I paraphrase the argument here. The main
contribution of his paper was to provide a quantitative statement of the
relation between the divergence of perturbation theory and the unstable
nature of the ground state in $\phi^3$ theory.

\begin{figure}[ht]
\centerline{\includegraphics[scale=0.4]{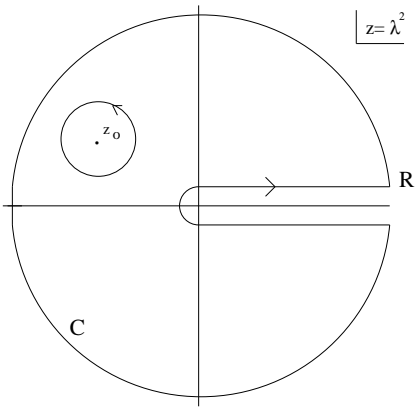}}
\caption{The complex $z=\lambda^2$ plane, showing the cut along the
positive $z$-axis.}
\label{f2}
\end{figure}

Motivated by the earlier results for $\phi^3$ theory (in 4 dimensions),
Arkady had the clever idea to consider  $\phi^3$ theory in $0+1$
dimensions, which is just quantum mechanics. Here it is natural to
consider the hamiltonian
\begin{eqnarray}
H=\frac{1}{2}\dot{\phi}^2+\frac{1}{2}m^2 \phi^2-\lambda \phi^3
\label{phi3ham}
\end{eqnarray}
\begin{figure}[ht]
\centerline{\includegraphics[scale=0.75]{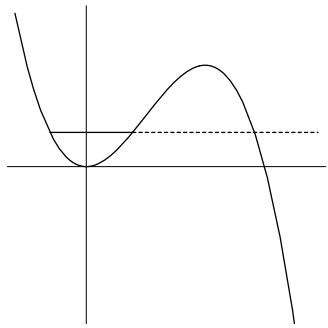}}
\caption{Unstable ground state for  $V(\phi)=\frac{1}{2}\phi^2-\lambda
\,\phi^3$}
\end{figure}
%\begin{figure}[th]
%\epsfxsize=5cm   %width of figure - will enlarge/reduce the figures
%\epsfbox{f1.eps}
%\figurebox{2cm}{3cm}{} %to have a box alone
%\centerline{\epsfxsize=3.5in\epsfbox{f1.eps}}  
%\caption{Unstable ground state for  $V(x)=x^2-\lambda \,x^3$}
%\end{figure}
and the ground state $|\psi\rangle$ such that $H|\psi
\rangle=E|\psi\rangle$. To make connection with the field theory results,
note that the two-point function
\begin{eqnarray}
i\, G(t)=\frac{\langle \psi |T\, \phi(t)\, \phi(0)\,
|\psi\rangle}{\langle\psi | \psi \rangle}
\label{green}
\end{eqnarray}
is related to the energy $E$ as
\begin{eqnarray}
i\, G(t=0)=\frac{1}{m}\, \frac{d\, E}{dm} =\frac{1}{m^2}\left(E-5
\lambda^2\,
\frac{d E}{d
\lambda^2}\right)
\label{ge}
\end{eqnarray}
where in the last step we have used the fact that, by dimensional reasoning,
the energy $E$ can be expressed as $E=m\,f(\frac{\lambda^2}{m^5})$.
Thus, if the perturbative expression for the two-point function diverges,
 the expression for the ground state energy, $E=E(\lambda^2)$, should
also diverge. One subtlety here is that the state $|\psi\rangle$
is clearly unstable. Arkady showed in an appendix \cite{arkady} how to
deal with this, by considering the adiabatic evolution of a stable state
into an unstable state. In particular this suggests that the expression
for $E=E(\lambda^2)$ must have a cut along the positive $\lambda^2$ axis,
as shown in Fig.~\ref{f2}, with an associated jump in the imaginary part
of $E$ across this cut.

Under the (important) assumption that there are no other cuts or poles in
the complex
$z=\lambda^2$ plane,  Cauchy's theorem implies that : 
\begin{eqnarray}
E(z_0)&=&\frac{1}{2\pi i}\oint_C dz\,
\frac{E(z)}{z-z_0}\nonumber\\\nonumber\\ &=&\frac{1}{\pi}\int_0^R dz\,
\frac{Im\, E(z)}{z-z_0}\nonumber\\\nonumber\\
 &=& \sum_{n=0}^\infty z_0^n\,
\left(\frac{1}{\pi}\int_0^R dz\,
\frac{Im\, E(z)}{z^{n+1}}\right)
\label{connection}
\end{eqnarray}
Thus, the perturbative expansion coefficients are explicitly related to the
moments of the imaginary part of the energy along the cut. Furthermore, it
is clear from (\ref{connection}) that at large $n$ (i.e., at large order in
perturbation theory), the dominant contribution comes from the behavior of
$Im E(z)$ as $z\to 0$. This observation is very important,
because the $z=\lambda^2\to 0$ limit is a semiclassical limit (note that
the barrier height goes like $\frac{1}{\lambda^2}$, and the barrier width
like
$\frac{1}{\lambda}$).  Hence, in this limit the imaginary part of the energy
may be estimated using semiclassical techniques, such as WKB. Simple
scaling shows that the leading WKB approximation for the imaginary part of
the energy has the form
\begin{eqnarray}
Im\, E(z)\sim \frac{a}{\sqrt{z}}\, e^{-b/z}\quad ,\quad z\to 0
\label{arkest}
\end{eqnarray}
where $a$ and $b>0$ are (calculable) constants. Note, of course, that this
expression is nonperturbative in $z=\lambda^2$.

\begin{figure}[ht]
\centerline{\includegraphics[scale=0.8]{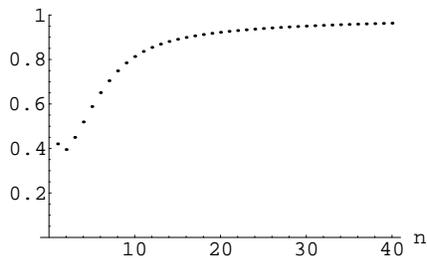}}
\caption{The ratio of the exact perturbative coefficients $c_n$ in
(\protect{\ref{cnexp}}) to the leading WKB expression from
(\protect{\ref{cnx3}}), as a function of the order
$n$.}
\label{f3}
\end{figure}

Now consider the perturbative expansion for the lowest energy
\begin{eqnarray}
E(\lambda^2)=\frac{1}{2}+\sum_{n=1} c_n \lambda^{2n}
\label{cnexp}
\end{eqnarray}
Inserting the WKB estimate (\ref{arkest}) into the dispersion relation
(\ref{connection}), we see that at large order the perturbation theory
coefficients should behave as 
\begin{eqnarray}
c_n\sim \frac{a}{\pi}\int_0^\infty dz\,
\frac{e^{-b/z}}{z^{n+3/2}}=\frac{a}{\pi}\,\frac{\Gamma(n+\frac{1}{2})}{b^{n+1/2}}
\label{arkdiv}
\end{eqnarray}
So, this argument suggests that the perturbative expansion (\ref{cnexp}) for
$E(\lambda^2)$ should be a divergent nonalternating series. Indeed, it is
straightforward to do this perturbative calculation to very high orders, and
to do the WKB calculation precisely, and one finds excellent
agreement \cite{yaris}. If the hamiltonian is rescaled as
$H=p^2+\frac{1}{4} x^2-\lambda x^3$ (this scaling makes the expansion
coefficients integers), then the leading growth rate for large $n$ is
\begin{eqnarray}
c_n\sim
-\frac{(60)^{n+1/2}}{(2\pi)^{3/2}}\,\,\Gamma(n+\frac{1}{2})
\left[1-\frac{169}{60(2n-1)}+O\left(\frac{1}{n^2}\right)\right]
\label{cnx3}
\end{eqnarray}
which agrees with Arkady's form (\ref{arkdiv}), and fits 
beautifully the growth rate of the actual expansion
coefficients \cite{yaris}. Indeed, the factorial growth of the
perturbative coefficients kicks in rather early, as illustrated in
Fig.~\ref{f3}.  

\noindent

The most important physics lessons from Arkady's paper
\cite{arkady} are :

(i) the divergence of perturbation theory is related to the possible
instability of the theory, at some phase of the coupling. 

(ii) there is a precise {\it quantitative} relation (\ref{connection})
between the large-order divergence of the perturbative coefficients and
nonperturbative  physics.

In more modern language, this divergence associated with instability and
tunneling is a divergence due to instantons. This idea has become a
cornerstone of quantum field theory \cite{abc}. However, since this time,
it has been found that in quantum field theory (as distinct from quantum
mechanics) there is yet another source of divergence in perturbation
theory -- this divergence is due to "renormalons", which arise
essentially because of the running of the coupling constant, and can be
related to special classes of diagrams \cite{thooft}. For recent
developments in this important subject, see the talks by M. Beneke, I.
Balitsky and E. Gardi in these Proceedings.

The connection between the large-order behavior of quantum mechanical
perturbation theory and WKB methods was developed independently, and
was probed in great depth, by Bender and Wu, who studied the quartic
anharmonic oscillator \cite{bw}. Bender and Wu developed recursion
techniques for efficiently generating very high orders of perturbation
theory, and compared these results with higher orders of the WKB
approximation. Their WKB analysis is a {\it tour de force}, and the
agreement with the large order perturbative coefficients is spectacular. 
\begin{figure}[ht]
\centerline{\includegraphics[scale=.9]{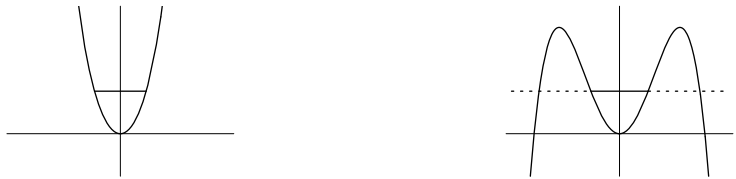}}
\caption{Stable ground state for the potential
$V=x^2+\lambda\, x^4$, on the left, and unstable ground state for the
potential $V=x^2-\lambda\, x^4$, on the right.}
\label{f4}
\end{figure}

The dependence of the ground state energy on the coupling $\lambda$ in the
$\lambda \phi^4$ case is different from the $\lambda \phi^3$ case.
As is clear from Fig. \ref{f4}, in the $\lambda\phi^4$ case the
instability arises when the coupling $\lambda$ changes sign from positive
to negative. Thus, the cut is expected along the {\it negative} real axis
in the complex $z=\lambda$ plane, which led Bender and Wu to the following
dispersion relation:
\begin{eqnarray}
c_n=\frac{1}{2\pi i}\int_{-\infty}^0 \frac{dz}{z^n} \lim_{\epsilon\to
0}\left[F(z+i\epsilon)-F(z-i\epsilon)\right]
\label{bwdispersion}
\end{eqnarray}
where $F(\lambda)=\frac{1}{\lambda}(E(\lambda)-\frac{1}{2})$ has one
subtraction. Here, Bender and Wu made use of some rigorous results
\cite{loeffel} concerning the analyticity behavior of $E(\lambda)$ in the
complex $\lambda$ plane : (i) $|E(\lambda)|\sim |\lambda|^{1/3}$ for large
$|\lambda|$, (ii) $E(\lambda)$ is analytic in the cut $\lambda$-plane,
with the cut along the negative real axis, and (iii) the expansion
$E(\lambda)\sim \frac{1}{2} +\sum_{n=1}^\infty c_n \lambda^n$ is asymptotic.

The dispersion relation (\ref{bwdispersion}) relates the perturbative
expansion coefficients for $F(\lambda)$ to the discontinuity of
$F(\lambda$) across the cut. Bender and Wu used high orders of WKB to
compute this nonperturbative imaginary part, thereby providing an extremely
precise connection between the large orders of perturbation theory and
semiclassical tunneling processes. They found that the expansion
coefficients are alternating and factorially divergent
\begin{eqnarray}
c_n=(-1)^{n+1}
\frac{3^n\,\sqrt{6}\,
\Gamma(n+\frac{1}{2}) }{\pi^{3/2}}\left[1-\frac{95}{72
n}+O\left(\frac{1}{n^2}\right)\right]
\label{cnx4}
\end{eqnarray}
Bender and Wu extended these results 
to the general anharmonic oscillator:
\begin{eqnarray}
\left(-\frac{d^2}{dx^2}+\frac{1}{4}x^2+\frac{\lambda}{2^N}\,
x^{2N}-E(\lambda)\right) \psi=0
\end{eqnarray}
Then the $k^{\rm th}$ energy level has an asymptotic series expansion
\begin{eqnarray}
E^{k,N}(\lambda)\sim k+\frac{1}{2}+\sum_{n=1}^\infty c_n^{k,N}
\,\lambda^n
\end{eqnarray}
where the perturbative expansion coefficients $c_n^{k,N}$ are related to
the lifetime of the $k^{\rm th}$ unstable level when the coupling $\lambda$
is negative:
\begin{eqnarray}
 c_n^{k,N}=
\frac{(-1)^{n+1}(N-1)2^k\Gamma(n N-n+k+\frac{1}{2})}{\pi^{3/2}k!2^n}
\left(\frac{\Gamma(\frac{2N}{N-1})}{\Gamma^2(\frac{N}{N-1})}
\right)^{nN-n+\frac{1}{2}} \hskip-15pt
\left[1+\dots\right]
\nonumber\\
\label{cnx2N}
\end{eqnarray}
Once again, these growth estimates are derived from WKB and fit the actual
perturbative coefficients with great precision \cite{bw}. 

%It is worth mentioning that the estimates (\ref{cnx3}), (\ref{cnx4}),
%(\ref{cnx2N}), can also be obtained from a simple scaling estimate due to
%B. Simon \cite{simonest}, who noted that at $n^{\rm th}$ order in
%Rayleigh-Schr\"odinger perturbation theory the contribution goes like
%\begin{eqnarray}
%\langle \psi | V\,\left(\frac{1}{H_0+E}\, V\right)^n\, |\psi\rangle
%\end{eqnarray}
%where $V$ is the perturbation, $H_0$ is the free harmonic oscillator
%hamiltonian, and $\psi$ is the unperturbed wavefunction. For example, for
%the ground state of the quartic anharmonic oscillator, at large $n$ this
%can be estimated as
%\begin{eqnarray}
%\langle \psi | V\,\left(\frac{1}{H_0+E}\, V\right)^n\, |\psi\rangle \sim
%\int dx\, e^{-x^2}\, x^4 \left(\frac{1}{x^2} \, x^4\right)^n \sim
%(n+\frac{3}{2})!
%\end{eqnarray}
%in agreement with (\ref{cnx4}). The other cases are similar.

An important distinction between the $\lambda\phi^3$ and $\lambda\phi^4$
quantum mechanical oscillators is that the perturbative series for the
energy eigenvalue is {\it nonalternating} in the $\lambda\phi^3$ case, and
{\it alternating} in the
$\lambda\phi^4$ case. This is directly related to the fact that the
$\lambda\phi^3$ case is inherently unstable (for any real $\lambda$),
while the
$\lambda\phi^4$ case is stable if $\lambda>0$, but unstable if
$\lambda<0$. Even though the quartic anharmonic oscillator with
$\lambda>0$ is completely stable, the perturbative expression for the
ground state energy is divergent, because the theory with $\lambda<0$ is
unstable.

It is interesting to note that there has been much recent
interest in certain nonhermitean hamiltonians, whose spectra appear to be
completely real, despite the nonhermicity \cite{stefan}. For example,
there is very strong numerical evidence \cite{bdx3,bwen} that the
hamiltonian 
\begin{eqnarray}
H=p^2+\frac{1}{4}x^2+i\epsilon x^3
\label{ix3}
\end{eqnarray}
has a completely real spectrum. For the massless case, $H=p^2+i\epsilon
x^3$, the reality of the spectrum has in fact been proved rigorously
\cite{dorey}. These results are nicely consistent with Arkady's analysis of
the $\lambda\phi^3$ quantum mechanical model, since an imaginary coupling
corresponds to $z=\lambda^2$ being real and negative. In this region it was
assumed that $E(z)$ is analytic (recall from Fig.~\ref{f2} that the cut is
along the {\it positive} $z$ axis), and the perturbative series is
divergent but alternating, and can be analyzed using various standard
(Pad\'e and Borel) techniques \cite{bdx3,bwen}, yielding excellent
agreement with numerical integration results.

A useful mathematical technique for dealing with
divergent series is Borel summation
\cite{hardy,bo,zinnbook,loeffel}. This method is best illustrated
by the paradigmatic case of the factorially divergent series. The series
\begin{eqnarray} 
f(g)\sim \sum_{n=0}(-1)^n\, n!\, g^{n}
\label{fact}
\end{eqnarray}
is clearly divergent, and is alternating if $g>0$. Using the standard
integral representation, $n!=\int_0^\infty ds\, e^{-s} s^n$, and formally
interchanging the summation and integration, we can write 
\begin{eqnarray}
f(g)\sim\frac{1}{g}\int_0^\infty ds\, {e^{-s/g}\over 1+s} 
\label{factint}
\end{eqnarray}
This integral representation is {\it defined} to be the Borel sum of the
divergent series in (\ref{fact}). The advantage of the integral is that it
is convergent for all $g>0$. To be more precise, all this actually
shows is that the integral in (\ref{factint})  has the same asymptotic
series expansion as  the divergent series in (\ref{fact}). In order for
this identification between the series and the Borel integral to be unique,
various further conditions must be satisfied \cite{hardy,bo,zinnbook}. In
some quantum mechanical examples it is possible to study these conditions
rigorously \cite{loeffel}, but unfortunately this is usually impossible in
realistic quantum field theories. This means we are often confined to
"experimental mathematics" when applying Borel techniques to perturbation
theory in QFT. Nevertheless, I prefer the attitude of  Heaviside:

\smallskip
\centerline{\fbox{\shortstack{\\``The series is divergent; therefore
we may be\\ able to do something with it''\\~ O. Heaviside, 1850 --
1925\\~}}}
\medskip

\noindent
to the (older) attitude of Abel:
\medskip

\centerline{\fbox{\shortstack{\\``Divergent series
are the invention of the devil, and it is shameful to\\ base on them any
demonstration whatsoever''\\~
N. H. Abel, 1828\\~}}}
\medskip

\noindent
Continuing the paradigm in (\ref{fact}), when $g<0$ the series
(\ref{fact}) becomes nonalternating. Then the same formal steps lead to
the following representation:
\begin{eqnarray}
f(-g)\sim\frac{1}{g}\int_0^\infty ds\,{e^{-s/g}\over 1-s} 
\end{eqnarray}
Clearly there is a problem here, as there is a pole on the contour of
integration, and so an ambiguity enters in the way one treats this pole.
This means that the nonalternating factorially divergent series in
(\ref{fact}), with $g<0$, is {\it not} Borel summable. For example, the
principal parts prescription leads to the following imaginary part for the
nonalternating series :
\begin{eqnarray}
Im f(-g)  = \frac{\pi}{g}\,\exp\left[-\frac{1}{g}\right]
\end{eqnarray}
This imaginary contribution is nonperturbative in the expansion parameter
$g$. It is not seen at any finite order in perturbation theory. However,
the imaginary part is inherently ambiguous in the absence of further
physical information beyond the series expansion (\ref{fact}) itself. 

Despite this ambiguity, it should be clear that the Borel approach provides
a natural formalism with which to analyze the problem of the divergence of
perturbation theory. It captures the essence of the connection with
nonperturbative tunneling, and associates such nonperturbative effects with
the unstable cases of the $\lambda \phi^3$ oscillator and the
$\lambda \phi^4$ oscillator with $\lambda<0$, for which the perturbative
series is indeed nonalternating and factorially divergent. 

Similar formal expressions exist for the Borel sum of $f(g)\sim\sum c_n
g^n$, if the  $c_n$ are not simply factorial as in
(\ref{fact}), but have the general form:
\begin{eqnarray}
c_n\sim\beta^n\, \Gamma(\gamma\, n
+\delta)
\label{cngeneral}
\end{eqnarray}
where $\beta$, $\gamma>0$ and $\delta$ are constants. Then the Borel sum
approximation is
\begin{eqnarray}
f(g)\sim \frac{1}{\gamma} \int_0^\infty \frac{ds}{s}
\left(\frac{1}{1+s}\right) \left(\frac{s}{\beta
g}\right)^{\delta/\gamma} \, \exp
\left[-\left(\frac{s}{\beta g}\right)^{1/\gamma}\right]
\label{genborel}
\end{eqnarray}
For $g<0$ the nonalternating series has an 
imaginary part:
\begin{eqnarray}
Im\,f(-g)\sim \frac{\pi}{\gamma} \left(\frac{1}{\beta
g}\right)^{\delta/\gamma} \, \exp \left[-\left(\frac{1}{\beta
g}\right)^{1/\gamma}\right]
\label{imborel}
\end{eqnarray}
In the next section we will use these relations in an explicit example.

Much more  could be said about the divergence of perturbation
theory, both in quantum mechanics and field theory. Lipatov 
\cite{lipatov}
generalized the instanton technique to scalar field theory, showing that
large orders of perturbation theory may be described by 
pseudoclassical solutions of the classical field equations, together with
quantum fluctuations. This approach built on instanton results of Langer
\cite{langer} in his classic study of metastability. Perturbation theory
for systems with degenerate minima was investigated in
\cite{degenerate}, and a new twist on the perturbative--nonperturbative
connection for this degenerate case is discussed in section 3 of this
talk. For further references, I refer the interested reader to the review
\cite{leguillou} of Le Guillou and Zinn-Justin as an excellent source.

\section{Effective actions, OPEs, and divergent
series}

As mentioned previously, it is extremely difficult, even in quantum
mechanics, to prove truly rigorous results concerning the divergence of
perturbation theory \cite{loeffel}; in quantum field theory we are even
more restricted when it comes to rigor. However, the study of effective
actions is an example in QFT where some rigorous results are possible.
This also makes connection with the subject of operator product
expansions, which is another subject to which Arkady has made seminal
contributions.

\begin{figure}[ht]
\centerline{\includegraphics[scale=0.25]{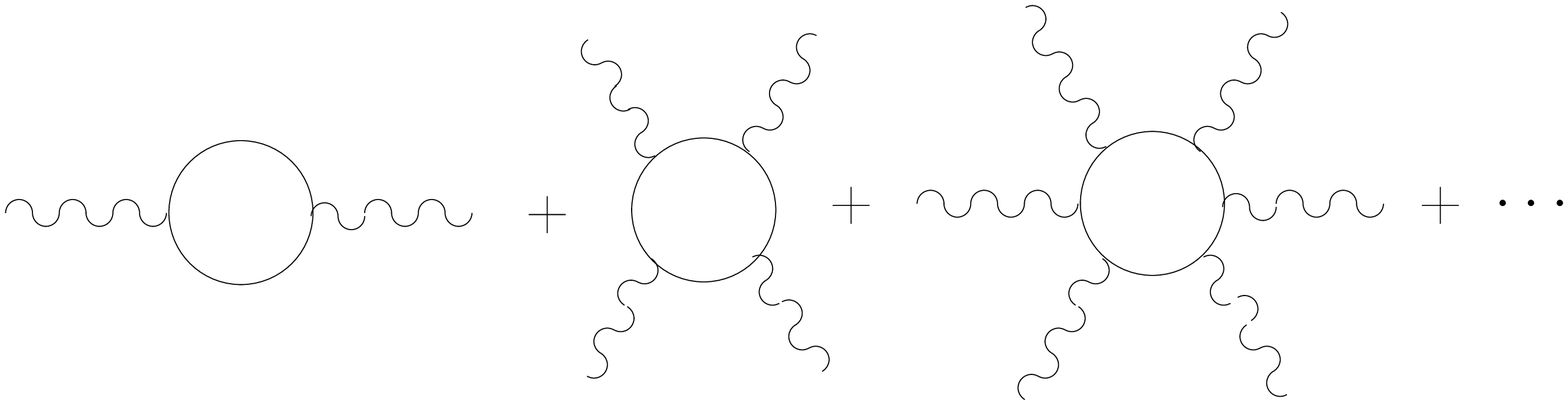}}
\caption{Perturbative expansion of the one-loop effective action}
\label{f5}
\end{figure}

For this talk, I consider the QED effective action,
which encodes nonlinear interactions due to quantum vacuum polarization
effects, such as light-by-light scattering. The effective action is defined
via the determinant that is obtained when the electron fields are
integrated out of the QED functional integral:
\begin{eqnarray}
S[A]=-\frac{i}{2} \log \det \left( D\hskip -6pt /
^2+m^2\right)
\end{eqnarray}
where $D\hskip -6pt /=\gamma^\mu D_\mu=\gamma^\mu (\partial_\mu-ieA_\mu)$ is
the Dirac operator in the classical gauge field background $A_\mu$.  The
effective action has a natural perturbative expansion in terms of the
electromagnetic coupling $e$, as represented in Fig.~\ref{f5}. 
Indeed, by
charge conjugation invariance (Furry's theorem), the expansion 
involves only even numbers of external photon lines, which means that the
perturbative series is actually a series in the fine structure constant
$\alpha=\frac{e^2}{4\pi}$. Another natural expansion is the "effective field
theory" expansion (or "large mass" expansion):
\begin{eqnarray}
S[A]=m^4\sum_n \, c_n\,\frac{O^{(n)}}{m^n}
\label{eff}
\end{eqnarray}
Here  $O^{(n)}$ represents gauge invariant and Lorentz invariant terms
constructed from the field strength $F_{\mu\nu}$, and having mass
dimension $n$. For example, at mass dimension $8$, we can have
$(F_{\mu\nu}F^{\mu\nu})^2$ or $(F_{\mu\nu}\tilde{F}^{\mu\nu})^2$, while at
mass dimension $10$ we could have a term $(\partial_\mu F_{\nu\rho}
\partial^\mu F^{\nu\rho})(F_{\alpha\beta}F^{\alpha\beta})$. As shown by
Arkady and his collaborators \cite{arkadyope,arkadyfs}, such an expansion
is related to the operator product expansion (OPE), such as that for
\begin{eqnarray}
\Pi_{\mu\nu}=(q_\mu q_\nu -q^2
g_{\mu\nu})\sum_n c_n(Q^2) \,\langle O_n\rangle
\label{ope}
\end{eqnarray}
In the special case where the classical background has constant field
strength $F_{\mu\nu}$, the perturbative and large mass expansions coincide.
This case of a constant background field was solved by  Euler and
Heisenberg \cite{euler} (see also Weisskopf \cite{weisskopf} and Schwinger
\cite{schwinger}), who obtained an exact  nonperturbative
expression for the effective action:
\begin{eqnarray}
S[A]&=&\frac{1}{8\pi^2}\int_0^\infty \frac{ds}{s^3}\, e^{-im^2 s}\left\{
(es)^2 |{\mathcal G}|\cot\left[es\left( \sqrt{{\mathcal F}^2+{\mathcal G}^2}+{\mathcal
F}\right)^{\frac{1}{2}}\right]\right.\nonumber\\
&&\times \left. \, \coth\left[es\left(
\sqrt{{\mathcal F}^2+{\mathcal G}^2}-{\mathcal
F}\right)^{\frac{1}{2}}\right]-1+\frac{2}{3}(es)^2 {\mathcal F}\right\}
\label{eh}
\end{eqnarray}
Here ${\mathcal F}=\frac{1}{4} F_{\mu\nu}F^{\mu\nu}$, and ${\mathcal G}=\frac{1}{4}
F_{\mu\nu}\tilde{F}^{\mu\nu}$, are the two Lorentz invariant combinations.
The $-1$ term in the integrand corresponds to the zero-field subtraction
of $S[0]$, while the last term, $\frac{2}{3}(es)^2 {\mathcal F}$,
corresponds to charge renormalization \cite{schwinger}. There are
several important physical  consequences of this result
\cite{euler,weisskopf,schwinger}. First, expanding to leading order in
the fields, we find the famous light-by-light term:
\begin{eqnarray}
S=\frac{2\alpha^2}{45 m^4}\int d^4x\,
\left[(\vec{E}^2-\vec{B}^2)^2+7(\vec{E}\cdot\vec{B})^2\right]+\dots
\label{lbl}
\end{eqnarray}
Second, for a constant electric field background, there is an imaginary part
\begin{eqnarray}
Im\, S = \frac{e^2 E^2}{8\pi^3}\sum_{k=1}^\infty
\frac{1}{k^2}\,
\exp\left[-\frac{k\, m^2\pi}{eE}\right]
\label{pp}
\end{eqnarray}
which gives the pair production rate due to vacuum polarization. 

\begin{figure}[ht]
\centerline{\includegraphics[scale=0.9]{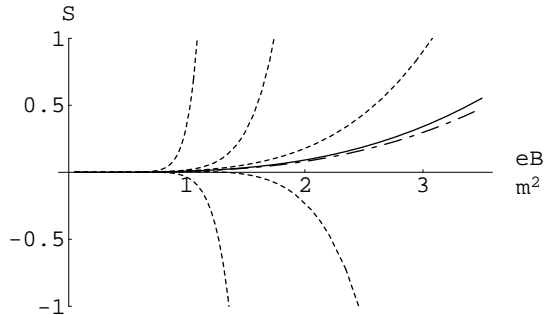}}
\caption{Comparison of the leading Borel expression
(\protect{\ref{leadingborel}}) [short-long-dash curve], as a function of
$\frac{eB}{m^2}$, with the exact expression (\protect{\ref{fullborel}})
[solid curve], and successive partial sums from the series
(\protect{\ref{ehmag}}) [short-dash curves]. The leading Borel expression
is much better than the series expressions for $\frac{eB}{m^2}\geq 1$.}
\label{f6}
\end{figure}

The Euler-Heisenberg result (\ref{eh}) provides an excellent example of the
application of Borel summation techniques, as we now review. Consider first
of all the case of a uniform magnetic field background of strength $B$.
Then the full perturbative expansion of the Euler-Heisenberg result is
\begin{eqnarray}
S=-\frac{e^2B^2}{2\pi^2}\sum_{n=0}^\infty {{\mathcal
B}_{2n+4}\over (2n+4)(2n+3)(2n+2)} \left(\frac{2eB}{m^2}\right)^{2n+2}
\label{ehmag}
\end{eqnarray} 
Viewed as a low energy effective action, the ``low energy" condition here
is simply that the characteristic energy scale for electrons
in the magnetic background, $\hbar\frac{eB}{mc}$, is much smaller than the
electron rest mass  scale $m c^2$. The expansion coefficients in the series
(\ref{ehmag}) involve the Bernoulli numbers ${\mathcal B}_{2n}$, which
alternate in sign and grow factorially in magnitude \cite{abramowitz}.
Thus, the series (\ref{ehmag}) is an alternating divergent series. In
fact, the expansion coefficients are:
\begin{eqnarray}
c_n&=&{2^{2n}\,{\mathcal
B}_{2n+4}\over (2n+4)(2n+3)(2n+2)} \nonumber\\
&=& (-1)^{n+1}
\,{\Gamma(2n+2)\over 8}\left[\frac{1}{\pi^{2n+4}}+\frac{1}{(2\pi)^{2n+4}}+
\frac{1}{(3\pi)^{2n+4}}+\dots\right]
\label{cnmag}
\end{eqnarray}
If we keep just the leading term in (\ref{cnmag}), then the
expansion coefficients are of the form in (\ref{cngeneral}), so that the
Borel prescription (\ref{genborel}) yields the leading Borel approximation
for $S$ as
\begin{eqnarray}
S_{\rm Borel} \sim {e^2B^2\over 4\pi^6} \int_0^\infty ds\,
{s\over 1+s^2/\pi^2} \,\exp\left[-{m^2 s\over eB}\right]
\label{leadingborel}
\end{eqnarray}
In Fig.~\ref{f6} this leading Borel approximation is compared with
successive terms from the perturbative expansion (\ref{ehmag}). Clearly the
Borel representation is far superior for larger values of the
expansion parameter $\frac{eB}{m^2}$.

Actually, in this Euler-Heisenberg case, we can do even better since we are
in the unusual situation of having the {\it exact} expression (\ref{cnmag})
for the perturbative coefficients, to all orders. Furthermore, the
subleading terms in (\ref{cnmag}) are also of the form (\ref{cngeneral}),
for which we can once again use the Borel prescription (\ref{genborel}).
Including these subleading terms, we find
\begin{eqnarray}
S= -{e^2B^2\over 8\pi^2}\int_0^\infty {ds\over s^2}
\left(\coth s-{1\over s}-{s\over 3} \right) \,\exp\left[-{m^2 s\over eB}
\right]
\label{fullborel}
\end{eqnarray}
where we have used the trigonometric expansion \cite{abramowitz}:
\begin{eqnarray}
\sum_{k=1}^\infty \frac{-2s^3}{k^2\pi^2
(s^2+k^2\pi^2)}=\coth s-\frac{1}{s}-\frac{s}{3}
\label{coth}
\end{eqnarray}
It is interesting to note that the correct renormalization subtractions
appear here.  The integral representation (\ref{fullborel}) is
precisely the Euler-Heisenberg expression (\ref{eh}), specialized to the
case of a purely magnetic field. Thus, the "proper-time" integral
representation (\ref{fullborel}) is the Borel sum of the divergent
perturbation series (\ref{ehmag}). Conversely, the divergent
 series (\ref{ehmag}) is the asymptotic expansion of the
nonperturbative Euler-Heisenberg result (\ref{eh}).

Now consider the case of a purely {\it electric} constant background, of
strength
$E$. Perturbatively, the only difference from the constant magnetic case is
that we replace $ B^2 \to - E^2$. This is because the only Lorentz
invariant combination is $(E^2-B^2)$ (clearly, $\vec{E}\cdot\vec{B}=0$ if
one or other of $E$ or $B$ is zero). Thus, the alternating series
in (\ref{ehmag}) becomes nonalternating, without changing the magnitude of
the expansion coefficients. Applying the Borel
dispersion relation (\ref{imborel}) leads to an imaginary part in exact
agreement with the nonperturbative pair-production result (\ref{pp}).

The divergence of the Euler-Heisenberg effective action is very
well known \cite{ioffe,og,chadha,pet,zhitnitsky}. This example is
analogous to Dyson's argument. The Euler-Heisenberg perturbative series
{\it cannot} be convergent, because if it were convergent, then for very
weak fields there would be no essential difference between the magnetic
or electric background. However, we know from  nonperturbative physics
that there {\it is} an inherent vacuum instability in the case of an
electric background, and that this leads to an exponentially small
imaginary part to the effective action, which corresponds to pair
production due to vacuum polarization. 

The phenomenon of
pair production from vacuum is fascinating, but is very difficult to observe
because the exponential suppression factor is exceedingly small for
realistic electric field strengths. The critical electric field at which
the exponent becomes of order 1 is $E_c=\frac{m^2 c^3}{e\hbar}\sim
10^{16}\,V/cm$, which is still several orders of magnitude beyond the peak
fields obtained in the most intense lasers. See \cite{ringwald} for a recent
discussion of the prospects for observing vacuum pair production using an
X-ray free-electron laser.

Even if one can produce such an intense background electric field, it is
clear that it will not be a uniform field. Thus, it is important to
ask how the Euler-Heisenberg analysis is modified when the background
field is inhomogeneous. This is a very difficult problem in general. The
most powerful approach is through semiclassical WKB techniques
\cite{brezin,popov}. However, here I turn this question around and ask what
this issue can tell us about the {\it series} expansion of the effective
action when the background is inhomogeneous \cite{ted}. If the field
strength is inhomogeneous, then the large mass expansion (\ref{eff}) of the
effective action involves many more terms at a given order, since we can
now include terms involving derivatives of $F_{\mu\nu}$. In fact
\cite{fleigner}, the number of terms appears to grow factorially fast: 1,
2, 7, 36, ... . But there are two obvious problems with quantifying this
divergence. First, the expansion is not a true series, for the
simple reason that at successive orders many completely new types of terms
appear. Second, to learn anything nonperturbative one needs to go to very
high orders in the derivative expansion, which is extremely difficult.

Fortunately, there is a special case in which  both these problems are
circumvented at once \cite{ted}. Consider the particular inhomogeneous
magnetic field pointing in the $z$-direction, but varying in the
$x$-direction as:
\begin{eqnarray}
\vec{B}(x)=\vec{B}\,{\rm sech}^2\left(x/\lambda\right)
\label{bsech}
\end{eqnarray}
where $\lambda$ is an arbitrary inhomogeneity scale. In such a background
the effective action can be computed in closed-form using nonperturbative
techniques \cite{dh1}. Also, since the inhomogeneity of the
background is encoded solely in the dependence on the scale $\lambda$, the
large mass expansion (\ref{eff}) can be expressed as a true series. That
is, at a given order in the derivative expansion, all terms with the same
number
$d$ of derivatives combine to give a single contribution
$\propto\frac{1}{(m\lambda)^d}$. Moreover, since this case is exactly
soluble, we have access to {\it all orders} of the derivative expansion.
Indeed, for the inhomogeneous magnetic background (\ref{bsech}), the
inverse mass expansion (\ref{eff}) becomes a double sum
\begin{eqnarray}
\hskip -15pt S =-\frac{m^4}{8\pi^{3/2}}\sum_{j=0}^\infty
\frac{1}{(m\lambda)^{2j}}
\sum_{k=2}^\infty
\left(\frac{2eB}{m^2}\right)^{2k}{\Gamma(2k+j)\Gamma(2k+ j -2){\mathcal
B}_{2k+2j} \over j!(2k)!\Gamma(2k+j+\frac{1}{2})}
\label{allorders}
\end{eqnarray}
with the two expansion parameters being the derivative expansion parameter,
$\frac{1}{m\lambda}$, and the perturbative expansion parameter,
$\frac{eB}{m^2}$. Note that the expansion coefficients are known to all
orders, and are relatively simple numbers, just involving the Bernoulli
numbers and factorial factors. It has been checked in \cite{dh1} that the
first few terms of this derivative expansion are in agreement with explicit
field theoretic calculations, specialized to this particular background.

Given the explicit series representation in (\ref{allorders}), we can
check that the series is divergent, but Borel summable. This can be done
in several ways. One can either fix the order $k$ of the perturbative
expansion in (\ref{allorders}) and show that the remaining sum is Borel
summable, or one can fix the order $j$ of the derivative expansion in
(\ref{allorders}) and show that the remaining sum is Borel summable. Or,
one can sum explicitly the $k$ sum, for each $j$, as an integral of a
hypergeometric function, and show that for various values of
$\frac{eB}{m^2}$, the remaining derivative expansion is divergent but
Borel summable. These arguments do not prove rigorously that the double
series is Borel summable, but give a strong numerical indication that this
is the case. It is interesting to note that an analogous double-sum
structure also appears in the renormalon-OPE analysis in the talk by E.
Gardi in these Proceedings.

It is also possible to compute the closed-form effective action for a time
dependent electric field pointing in the $z$-direction:
\begin{eqnarray}
\vec{E}(t)=\vec{E}\,{\rm sech}^2\left(t/\tau\right)
\label{esech}
\end{eqnarray}
where $\tau$ characterizes the temporal inhomogeneity scale. A short-cut to
the answer is to note that we can simply make the replacements, $B^2\to
-E^2$, and $\lambda^2\to -\tau^2$, in the magnetic case result
(\ref{allorders}). In particular this has the consequence that the
alternating divergent series of the magnetic case becomes a non-alternating
divergent series, just as was found in the Euler-Heisenberg constant-field
case. Fixing the order $j$ of the derivative expansion, the
expansion coefficients behave for large
$k$ (with $j$ fixed) as
\begin{eqnarray}  
\hskip -15pt c_k^{(j)}=\frac{(-1)^{j+k}\Gamma(2k+j)
\Gamma(2k+j+2){\mathcal B}_{2k+2j+2}} {\Gamma (2k+3)\Gamma
(2k+j+\frac{5}{2})}\sim  2 
{\Gamma(2k+3j-\frac{1}{2})\over (2\pi)^{2j+2k+2}}
\label{dc}
\end{eqnarray}
Note that these coefficients are non-alternating and grow factorially
with $2k$, as in the form of (\ref{cngeneral}). Applying the Borel
dispersion formula (\ref{imborel}) gives
\begin{eqnarray} 
 Im S^{(j)}\sim {m^4\over 8\pi^3}
\left(\frac{eE}{m^2}\right)^{5/2}\, \exp\left[-\frac{m^2\pi}{eE}\right]
\, \frac{1}{j!}\, \left({m^4 \pi\over 4 \tau^2 e^3 E^3}\right)^j
\label{ej}
\end{eqnarray} 
Remarkably, this form can be resummed in $j$, yielding  a leading
exponential
\begin{eqnarray} 
 Im S\sim {m^4\over 8\pi^3}
\left(\frac{eE}{m^2}\right)^{5/2}\, \exp\left[-\frac{m^2\pi}{eE}
\left\{1-\frac{1}{4} \left(\frac{m}{eE\tau}\right)^2\right\}\right] 
\label{eresum}
\end{eqnarray}
We recognize the first term in the exponent as the leading exponent in
the constant field result (\ref{pp}). Thus, the second term may be viewed
as the leading {\it exponential} correction to the constant-field answer
(\ref{pp}). This is exactly what we set out to find, and we see that it
arose through the divergence of the derivative expansion. I stress that this
exponential correction is not accessible from low orders of the derivative
expansion. This gives a Dyson-like argument that the derivative expansion
{\it must} be divergent, since if it were not divergent, there would be no
essential difference between the electric and magnetic cases, and there
would be no correction to the exponent of the imaginary part of the
effective action. However, we know, for example from WKB, that there
{\it is} such a correction, and so the derivative expansion must be
divergent.

In fact, the situation is even more interesting than the result
(\ref{eresum}) suggests. We could instead have considered
doing the Borel resummation for the $j$ summations, at each fixed $k$.
Then for large $j$, the coefficients go as 
\begin{eqnarray}  
\hskip -15pt c_j^{(k)}=(-1)^{j+k}\frac{\Gamma(j+2k)
\Gamma(j+2k-2){\mathcal B}_{2k+2j}} {\Gamma (j+1)\Gamma
(j+2k+\frac{1}{2})}\sim  2^{\frac{9}{2}-2k} 
{\Gamma(2j+4k-\frac{5}{2})\over (2\pi)^{2j+2k}}
\label{dce}
\end{eqnarray}
which once again are non-alternating and factorially growing. Applying the Borel dispersion
formula (\ref{imborel}) gives
\begin{eqnarray} 
 Im S^{(k)}\sim {m^{3/2}\over 4\pi^3 \tau^{5/2}}\, {(2\pi
eE\tau^2)^{2k}
\over (2k)!} \, e^{-2\pi m\tau} 
\label{kp}
\end{eqnarray} 
Once again, this leading form can be resummed, yielding
\begin{eqnarray} 
 Im S\sim {m^{3/2}\over 8\pi^3 \tau^{5/2}}\, 
\exp\left[-2\pi m\tau\left(1-\frac{eE\tau}{m}\right)\right]
\label{presum}
\end{eqnarray}
Note that this leading exponential form of the imaginary part is different
from that obtained in (\ref{eresum}), and moreover, it is different 
from the constant-field case (\ref{pp}). The resolution of this puzzle
is that there are two competing leading exponential behaviours
buried in the double sum (\ref{allorders}), and the question of which one
dominates depends crucially on the relative magnitudes of the two
expansion parameters, the derivative expansion parameter, $\frac{1}{m\tau}$,
and the perturbative expansion parameter, $\frac{eE}{m^2}$. Another
important quantity is their {\it ratio}, since this sets the scale of the
corresponding gauge field:
\begin{eqnarray}
\frac{A(t)}{m}=\frac{eE\tau\, {\rm tanh}(t/\tau)}{m}
\sim \frac{eE\tau}{m} ={eE/m^2\over 1/(m\tau)}
\label{scales}
\end{eqnarray}
Thus, it is natural to define a ``nonperturbative'' regime, in which
$\frac{eE\tau}{m}\gg 1$. Then $m\tau\gg\frac{m^2}{eE}$, so that the
dominant exponential factor is $\exp[-\frac{m^2}{eE}]\gg\exp[-2\pi
m\tau]$. In this regime, the leading imaginary contribution to the
effective action is given by the expression (\ref{eresum}), and we note
that it is indeed nonperturbative in form, and the correction in the
exponent is in terms of the small parameter $\frac{m}{eE\tau}\ll 1$.
On the other hand, in the ``perturbative'' regime,
where $\frac{eE\tau}{m}\ll 1$, this means that $m\tau\ll\frac{m^2}{eE}$, so
that the dominant exponential factor is
$\exp[-2\pi m\tau]\gg\exp[-\frac{m^2}{eE}]$. In this regime, the leading imaginary contribution to the
effective action is given by the expression (\ref{presum}), and is in fact
perturbative in nature, despite its exponential form.

These results are completely consistent with the WKB approach developed by
Br\'ezin and Itzykson \cite{brezin} and Popov \cite{popov}. For a
time-dependent electric background $E(t)=\dot{A}_z(t)$, in the
$z$-direction, the WKB expression for the imaginary part of the effective
action is:
\begin{eqnarray}
Im S\sim \int d^3k\, \exp[-\pi \Omega]
\label{wkb}
\end{eqnarray}
where $\Omega =\frac{2i}{\pi}\int_{\rm tp}
\sqrt{m^2+k_\perp^2+(k_z-eA_z(t))^2}$. Applying this  WKB analysis
to the (exactly soluble) case $E(t)=E\,{\rm sech}^2(t/\tau)$, 
one  obtains precisely the leading results
(\ref{eresum}) or (\ref{presum}), depending on whether we are in the
``non-perturbative'' $\frac{eE\tau}{m}\gg 1$, or ``perturbative''
$\frac{eE\tau}{m}\ll 1$ regime. This serves as a useful cross-check of the
somewhat formal Borel analysis.

\section{Perturbative -- nonperturbative duality in QES systems}

In this last section I discuss some recent results \cite{ds} concerning a
new type of perturbative -- nonperturbative connection that has been
found in certain special quantum mechanical systems. We benefited from
discussing these systems with Arkady, and I hope he enjoys the results!

Quasi-exactly solvable (QES) systems are those for which some finite
portion of the energy spectrum can be found exactly using algebraic means
\cite{itep}. A positive integer parameter $J$ characterizes the `size' of
this exact portion of the spectrum. Two simple examples are : $
V=x^6-(4J-1)x^2$, and $V=\sinh^2 x-(2J-1)\cosh x$. For a QES system it is
possible to define a quadractic form, $H=\sum_{a,b} c_{ab} J_a\, J_b+\sum_a
d_a \, J_a$, in terms of  $sl(2)$ generators of spin $J$, such that the
eigenvalues of the algebraic matrix $H$ are the QES
eigenvalues of the original system. It is interesting that algebraic
hamiltonians of this form are widely used in the study of tunneling
phenomena in single-molecule magnets \cite{magnet}.

In \cite{bdm}, the large $J$ limit of QES systems was identified as a
semiclassical limit useful for studying the {\it top} of the quasi-exact
spectrum. It was found that remarkable factorizations reduce the
semiclassical calculation to simple integrals, leading to a straightforward
asymptotic series representation for the {\it highest} QES energy
eigenvalue. The notion of energy-reflection (ER) symmetry was introduced
and analyzed in
\cite{st2}: for certain QES systems the QES portion of the spectrum is
symmetric under the energy reflection $E\leftrightarrow - E$. This means
that for a system with ER symmetry, there is a precise connection between
the top of the QES spectrum and the bottom of the spectrum. Coupled with the
semiclassical large $J$ limit,  the ER symmetry therefore relates
semiclassical (nonperturbative) methods with perturbative methods
\cite{st2}. In this section I discuss a class of {\it periodic} QES
potentials for which the ER symmetry is in fact the fixed point
(self-dual point) of a more general duality transformation. The duality
between weak coupling and semiclassical expansions applies not just to
the asymptotic series for the locations of the bands and gaps, but also
to the exponentially small widths of bands and gaps. 

\begin{figure}[ht]
\centerline{\includegraphics[scale=0.75]{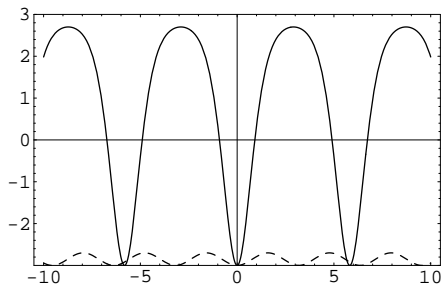}}
\caption{The potential energy in (\protect{\ref{lame}})
as a function of $\phi$, for $J=2$. The solid curve has
elliptic parameter $\nu =0.95$, for which the period is $2K(0.95)\approx
5.82$. The dashed curve has $\nu = 0.05$, for which the
period is $2K(0.05)\approx 3.18$. Note how different
the two potentials are; and yet, their spectra are related by the duality
transformation (\protect{\ref{duality}}).}
\label{f7}
\end{figure}

Consider the quasi-exactly solvable (QES) Lam\'e equation \cite{ww}:
\begin{eqnarray}
\left\{ -\frac{d^2}{d\phi^2} +J(J+1)\,\nu\,{\rm sn}^2(\phi|\nu )- 
\frac{1}{2}J(J+1)\right\}\Psi (\phi) = E\, \Psi (\phi)\,.
\label{lame}
\end{eqnarray}
Here ${\rm sn}(\phi|\nu )$ is the doubly-periodic Jacobi elliptic  function
\cite{ww,abramowitz}, the coordinate $\phi\in R^1$, and $E$ denotes
the energy eigenvalue. The real elliptic parameter $\nu$ lies in the range
$0\leq\nu\leq 1$.  The potential in (\ref{lame}) has period $2K(\nu)$, where 
$K(\nu)$ is the elliptic quarter period.
The parameter
$\nu$ controls the period of the potential, as well as its strength: see
Fig.~\ref{f7}. As
$\nu\to 1$, the period $2K(\nu)$ diverges logarithmically, $2K(\nu)\sim \ln\,
(\frac{16}{1-\nu})$, while as $\nu\to 0$, the period tends to a nonzero
constant: $2K(\nu)\to \pi$. In the Lam\'e equation (\ref{lame}), the
parameter $J$ is a positive integer (for non-integer $J$, the problem is not
QES). This parameter $J$ controls the depth of the wells of the potential;
 the constant subtraction $-\frac{1}{2}J(J+1)$ will
become clear below. 

It is a classic result that the Lam\'e equation (\ref{lame}) has bounded
solutions
$\Psi(\phi)$ with an energy spectrum consisting of exactly
$J$ bands, plus a continuum band \cite{ww}. It is the simplest example of a
``finite-gap"  model, there being just a finite number, $J$, of ``gaps" in
the spectrum. This should be contrasted with the fact that a generic
periodic potential has an {\it infinite} sequence of gaps in its spectrum
\cite{magnus}. We label the band edge energies by
$E_l$, with $l=1,2,\dots, (2J+1)$. Thus, the energy regions, $E_{2l-1}\leq E\leq
E_{2l}$, and $E\geq E_{2l+1}$, are the allowed bands, while
the regions, $E_{2l}< E < E_{2l+1}$, and $E< E_1$, are the gaps.

Another important classic result \cite{iachello,ward} concerning
the Lam\'e model (\ref{lame}) is that the band edge energies $E_l$, for
$l=1,\dots, 2J+1$, are simply the eigenvalues of the
finite dimensional $(2J+1)\times(2J+1)$ matrix
\begin{eqnarray}
H=J_x^2+\nu J_y^2 -\frac{1}{2} J(J+1)\, {\bf I}
\label{ham}
\end{eqnarray}
where $J_x$ and $J_y$ are $su(2)$ generators in a spin $J$ representation
and ${\bf I}$ is the unit matrix.
Thus the Lam\'e band edge spectrum is {\it algebraic}, requiring only the
finding of the eigenvalues of the finite dimensional matrix $H$ in
(\ref{ham}). For example, for $J=1$ and $J=2$, the eigenvalues of $H$ are:
\begin{eqnarray}
J=1\quad :\quad E_1&=&-1+\nu \, ,\nonumber\\
E_2&=& 0\,,\nonumber\\ 
E_3&=& \nu\,; 
\label{egj1}\\[2mm]
J=2\quad :\quad 
E_1&=&-1 + 2\,\nu - 2\,{\sqrt{1 - \nu +\nu^2}}\,,\nonumber \\
E_2&=& -2 + \nu\,,\nonumber \\
E_3&=& -2 + 4\,\nu  \,,\nonumber \\
E_4&=&1 + \nu\,, \nonumber \\
E_5&=&-1 + 2\,\nu + 2\,{\sqrt{1 - \nu + \nu^2}}\,.
\label{egs}
\end{eqnarray}

\begin{figure}[ht]
\centerline{\includegraphics[scale=0.8]{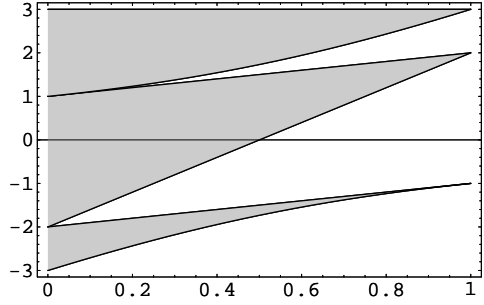}}
\caption{The energy bands (\protect{\ref{egs}}) for the $J=2$  Lam\'e
system (\ref{lame}), as a function of the elliptic parameter $\nu$. The
shaded areas on this plot are the bands, while the unshaded areas are the
gaps. The top band actually continues up to
$E\to\infty$.}
\label{f8}
\end{figure}
The spectrum of the Lam\'e system (\ref{lame}) has a special duality:
\begin{eqnarray}
E[\nu]=-E[1-\nu]
\label{duality}
\end{eqnarray}
That is, the spectrum of the Lam\'e system (\ref{lame}), with elliptic
parameter $\nu$, is the energy reflection of the spectrum of the Lam\'e
system with the dual elliptic parameter $1-\nu$. In particular, for the band
edge energies, $E_l$, which are the eigenvalues of the finite dimensional
matrix $H$ in (\ref{ham}), this means
\begin{eqnarray}
E_l[\nu]=-E_{2J+2-l}\,[1-\nu]\qquad , \quad l=1,2,\dots, 2J+1
\label{hduality}
\end{eqnarray}
This duality can be seen directly in the eigenvalues of the $J=1$ and $J=2$
examples in (\ref{egj1}) and (\ref{egs}). The proof of the duality
(\ref{hduality}) for the band edge energies is a trivial consequence of
the algebraic realization (\ref{ham}), since
\begin{eqnarray}
J_x^2+\nu J_y^2 -\frac{1}{2} J(J+1)\, {\bf I}
=  -\left[
\left(J_z^2+(1-\nu)J_y^2\right)-\frac{1}{2} J(J+1)\,{\bf I}\right]
\label{proof}
\end{eqnarray}
Noting that $[J_z^2+(1-\nu)J_y^2]$ has the same eigenvalues as
$[J_x^2+(1-\nu)J_y^2]$, the duality result (\ref{hduality}) follows. It is
instructive to see this duality in graphical form, in
Fig.~\ref{f8}, which shows the band spectra as a function of $\nu$. The
transformation $\nu\to 1-\nu$,  with the energy reflection $E\to
-E$, interchanges the shaded regions (bands) with the unshaded regions
(gaps). The fixed point, $\nu=\frac{1}{2}$, is the ``self-dual" point,
where the system maps onto itself; here the energy spectrum has an exact
energy reflection (ER) symmetry. 

In fact, the duality relation (\ref{duality}) applies to the entire
spectrum, not just the band edges (\ref{hduality}). This is a 
consequence of Jacobi's imaginary transformation \cite{abramowitz}. Making
the coordinate transformation
\begin{eqnarray}
\phi^\prime=i\left(\phi-K-i\, K^\prime\right)\,,
\label{rotation}
\end{eqnarray}
the Lam\'e
equation (\ref{lame}) transforms into
\begin{eqnarray}
\left\{ -\frac{d^2}{d\phi^{\prime 2}} +J(J+1)\,(1-\nu)\,
{\rm sn}^2(\phi^\prime|1-\nu)- 
\frac{1}{2}J(J+1)\right\}\Psi (\phi^\prime) =- E\, \Psi
(\phi^\prime)
\nonumber\\
\label{lameprime}
\end{eqnarray}
So solutions of  (\ref{lame}) are mapped to solutions of
the dual equation (\ref{lameprime}), with $\nu\to 1-\nu$, and with a sign
reflected energy eigenvalue: $E\to -E$. 

\begin{figure}[ht]
\centerline{\includegraphics[scale=1.1]{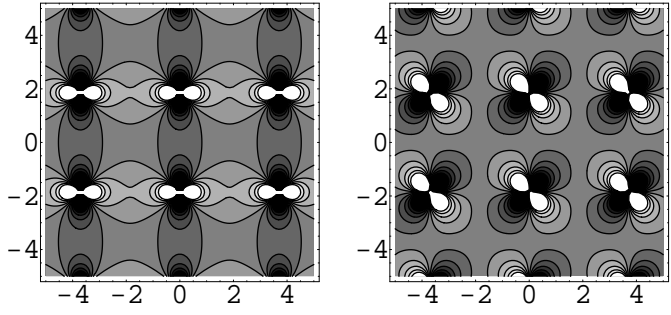}}
\caption{Contour plots in the $\phi$ plane of the real (left) and
imaginary (right) parts of the potential function $sn^2(\phi |\nu)$, for
$\nu=\frac{1}{2}$. Note that the potential is real along the real $\phi$
axis, and along the line $Re(\phi)=K(\frac{1}{2})\approx 1.85$ (as well
as the periodic displacements). This corresponds to the rotation 
(\protect{\ref{rotation}}).}
\label{f9}
\end{figure}

To see why bands and gaps are interchanged under our duality transformation
(\ref{rotation}), recall that the independent solutions of the original
Lam\'e equation (\ref{lame}) can be written  as products of theta functions
\cite{ww}, and under the change of variables (\ref{rotation}), these theta
functions map into the same theta functions, but with dual elliptic
parameter. However, they map from bounded to unbounded solutions (and vice
versa), because of the ``i" factor appearing in (\ref{rotation}). Thus, the
bands and gaps become interchanged. 

As an aside, I mention that the Lam\'e system plays a distinguished role
in the theory of $su(2)$ BPS monopoles. For example, Ward has shown
\cite{ward} that the Lam\'e equation factorizes, using the Nahm
equations (which are fundamental to the construction of monopole
solutions)
\begin{eqnarray}
\frac{d T_j}{ds}=\mp \frac{i}{2}\epsilon_{jkl}\, [T_k, T_l ]
\label{nahm}
\end{eqnarray}
To see this, define the quaternionic operators
\begin{eqnarray}
\Delta_\pm={\bf 1}_{2n}\frac{d}{ds}\pm
T_j(s)\otimes \sigma_j
\label{quat}
\end{eqnarray}
where $T_j(s)$, for $j=1,2,3$, take values in an $n$-dimensional
represenation of the Lie algebra $su(2)$. Then
\begin{eqnarray}
\Delta_\pm \Delta_\mp =\left({\bf 1}_n\frac{d^2}{ds^2}-T_jT_j\right) \otimes
{\bf 1}_2+\left(\mp \frac{d T_j}{ds}-\frac{i}{2}\epsilon_{jkl}\, 
[T_k, T_l ]\right)
\otimes \sigma_j
\label{nahmfact}
\end{eqnarray}
and this combination is real if $T_j$ satisfy the Nahm equations
(\ref{nahm}). Furthermore, using the solution
$T_1=-\sqrt{\nu}\, {\rm sn}(s) \, t_1$, $
T_2=i\sqrt{\nu}\, {\rm cn}(s) \, t_2$, $
T_3=i\, {\rm dn}(s) \, t_3$, 
where $t_j$ are $su(2)$ generators, it follows from properties of the
elliptic functions that
\begin{eqnarray}
T_j T_j=(t_1^2+t_2^2+t_3^2)\,\nu\, {\rm sn}^2(s)-(\nu\, t_2^2+t_3^2)
\end{eqnarray}
Thus, (\ref{nahmfact}) provides a factorization of the Lam\'e operator.
Subsequently, Sutcliffe showed \cite{sutcliffe} that the spectral curve for
the Nahm data for a charge $n=2j+1$ $su(2)$ monopole is related to  a
$j$-gap Lam\'e operator, and corresponds physically to $2j+1$ monopoles
aligned along an axis.

Returning to the perturbative--nonperturbative duality (\ref{duality}),
I first discuss how this operates for the {\it locations} of the bands and
gaps in the Lam\'e spectrum. The location of a low-lying band can be
calculated using perturbative methods, while the location of a high-lying
gap can be calculated using semiclassical methods. The exact duality
(\ref{duality}) between the top and bottom of the spectrum provides an
explicit mapping between these two sectors. Defining
$\kappa =\sqrt{J(J+1)}$, we see that $1/\kappa$ is the weak coupling
constant of the perturbative expansion. Simultaneously, $1/\kappa$ plays
the role of $\hbar$ in the quasiclassical expansion.

In the limit $J\to\infty$, the width of the lowest band becomes very
narrow, so it makes sense to estimate the ``location" of the band. In fact,
the width  shrinks exponentially fast, so we can estimate the location of
the band to within exponential accuracy using elementary perturbation
theory. A straightforward calculation \cite{ds} shows that the lowest
energy level is
\begin{eqnarray}
E_0 = -\frac{1}{2}\, \kappa^2 
\left[1-\frac{2\sqrt{\nu}}{\kappa}
+\frac{\nu+1}{2\, \kappa^2}+
\frac{(1-4
\nu+\nu^2)}{8\,\sqrt{\nu}\,\kappa^3} +O\left(\frac{1}{\kappa^4}\right)\right]\,.
\label{e0pt}
\end{eqnarray}
We now consider the semiclassical evaluation of the location of the
highest gap, in the limit $J\to\infty$. First, note that for a given
$\nu$, as
$J\to\infty$ the highest gap lies {\it above} the top of the potential.
Thus, the turning points lie off the real $\phi$ axis. For a periodic
potential the gap edges occur when the discriminant \cite{magnus} 
takes values $\pm 1$. By WKB, the discriminant is
\begin{eqnarray}
\Delta(E)=\cos \left( \frac{1}{\hbar}\,\,  \sum_{n=0}^\infty \, \hbar^n\,
S_n\, (P)\right)\,,
\label{disc}
\end{eqnarray}
where $P$ is the period, and  $S_n(x)$ are the standard 
WKB functions which can be generated to any order by a  simple
recursion formula \cite{bo}. The $J^{\rm th}$ gap occurs when the
argument of the cosine in the discriminant (\ref{disc}) is $J\pi$:
\begin{eqnarray}
\kappa\, \, \sum_{n=0}^\infty \,
\frac{1}{\kappa^n }  \, {S_n\, (2K)} = J\,\pi= 
\pi\, \kappa\left(1-\frac{1}{2\,
\kappa}+\frac{1}{8\,\kappa^2}-\frac{1}{128\,\kappa^4} +
\dots \right)
\label{jjj}
\end{eqnarray}
where $2K$ is the period of the Lam\'e potential, and where on the right-hand side
we have expressed $J$ in terms of the effective semiclassical  expansion parameter
$1/\kappa$. This relation (\ref{jjj}) can be used to find an expansion for the
energy of the $J^{\rm th}$ gap by expanding
\begin{eqnarray}
E=\frac{1}{2}\,\kappa^2 +\sum_{\ell=1}^\infty\, 
\frac{\varepsilon_\ell}{\kappa^{\ell-2}}\,.
\label{eexp}
\end{eqnarray}
The expansion coefficients $\varepsilon_l$ are fixed by identifying terms on both
sides of the expansion in (\ref{jjj}). A straightforward calculation
\cite{ds} leads to
\begin{eqnarray}
E=\frac{1}{2}\,\kappa^2
-\kappa\,
\sqrt{1-\nu}
+\frac{2-\nu}{4}+\frac{(-2+2\nu+\nu^2)}{16\,\kappa\,\sqrt{1-\nu}}
+\dots\,.
\label{wkbexp}
\end{eqnarray}
Comparing with the perturbative expansion (\ref{e0pt}) we see that the 
semiclassical expansion (\ref{wkbexp}) is indeed the dual of the
perturbative expansion (\ref{e0pt}), under the duality transformation
$\nu\to 1-\nu$ and $E\to -E$.

This illustrates the perturbative -- nonperturbative duality for the {\it
locations} of bands and gaps. However, it is even more interesting to
consider this duality for the {\it widths} of bands and gaps, because the
calculations of widths are sensitive to exponentially small contributions
which are neglected in the calculations of the locations. 

The width of a low-lying band can be computed in a number of ways. First,
since the band edge energies are given by the eigenvalues of the finite
dimensional matrix $H$ in (\ref{ham}), the most direct way to evaluate the
width of the lowest-lying band is to take the difference of the two smallest
eigenvalues of $H$. This leads \cite{dr} to the exact leading behavior, in
the limit
$\nu\to 1$, of the width of the lowest band, for any $J$ :
\begin{eqnarray}
\Delta E^{\rm algebraic}_{\rm band} = {8J\, \Gamma(J+1/2)\over 4^J
\sqrt{\pi}\,
\Gamma(J)} \, (1-\nu)^J\, \left(1+\frac{J-1}{2} (1-\nu)+\dots\right)
\label{exact}
\end{eqnarray}
This clearly shows the exponentially narrow character of the lowest band.

In the instanton approximation, tunneling is suppressed because the barrier
height is much greater than the ground state energy of any given isolated
``atomic" well. The instanton calculation for the Lam\'e
potential can be done in closed form \cite{dr}, leading in the large
$J$ and $\nu\to 1$ limit to  
\begin{eqnarray}
\Delta  E^{\rm instanton}_{\rm band}\sim \frac{8 J^{3/2}}{\sqrt{\pi}\, 4^J}
\,\,
\left(1-\nu\right)^J \left[ 1+\frac{J-1}{2}(1-\nu)+\dots \right]\,,
\label{largej}
\end{eqnarray} 
which agrees perfectly with the large $J$ limit  of
the exact algebraic result (\ref{exact}). Thus, this example gives an
{\it analytic} confirmation that the instanton approximation gives the
correct leading large $J$ behavior of the width of the lowest band, as
$\nu\to 1$.

Having computed the width of the lowest band by several different
techniques, both exact and nonperturbative, we now turn to a perturbative
evaluation of the width of the highest gap. First, taking the 
difference of the two largest eigenvalues of the finite
dimensional matrix $H$ in (\ref{ham}), it is straightforward to show 
that as $\nu\to 0$, for any $J$, this difference gives 
\begin{eqnarray}
\Delta E^{\rm algebraic}_{\rm gap} = {8J\, \Gamma(J+1/2)\over 4^J
\sqrt{\pi}\,
\Gamma(J)} \, \nu^J \left(1+\frac{J-1}{2}\, \nu+\dots\right)\,,
\label{exacttop}
\end{eqnarray}
which is the same as the algebraic expression (\ref{exact}) for the width
of the lowest band, with the duality replacement $\nu\to 1-\nu$. But it is
more interesting to try to find this result from perturbation
theory. From (\ref{exacttop}) we see that the width of the highest
gap is of $J^{\rm th}$ order in perturbation theory. So, to compare with
the semi-classical (large $J$) results for the width of the lowest band,
we see that we will have to be able to go to very high orders in
perturbation theory. This provides a novel, and very direct,
illustration of the connection between nonperturbative physics and high
orders of perturbation theory.

It is generally very difficult to go to high orders in perturbation theory,
even in quantum mechanics. For the Lam\'e system (\ref{lame}) we can
exploit the algebraic relation to the finite-dimensional spectral problem
(\ref{ham}). However, since $H$ in (\ref{ham}) is a $(2J+1)\times (2J+1)$
matrix, the large $J$ limit is still non-trivial. Nevertheless, the high
degree of symmetry in the Lam\'e system means that the perturbative
calculation can be done to arbitrarily high order \cite{ds}. The result is
that the splitting between the two highest eigenvalues arises at the $J^{\rm
th}$ order in perturbation theory, and is given by
\begin{eqnarray}
\Delta E^{\rm pert. theory}_{\rm gap}=\frac{8}{4^{2J}} \,\,
\frac{(2J)!}{\left[(J-1)!\right]^2}\,\nu^J= \frac{8 J\, \Gamma(J+1/2)}{4^J \sqrt{\pi}\,
\Gamma(J)}\, \nu^J
\label{pt}
\end{eqnarray}
This is in complete agreement with  (\ref{exacttop}), and by duality
agrees also with the nonperturbative results for the width (\ref{exact})
of the lowest band. 

\section{Conclusions}

To conclude, there are many examples in physics where there are divergences
in perturbation theory which can be associated with a potential instability
of the system, thereby providing an explicit bridge between the
nonperturbative and perturbative regimes. While this is not the only source
of divergence, it is an important one which involves much fascinating
physics. Arkady has made many important advances in this subject. On this
occasion it is especially appropriate to give him the last word: 
\vskip 1mm

\centerline{\fbox{\shortstack{``The majority of nontrivial theories 
are seemingly unstable at some\\
phase of the coupling constant, which leads to the asymptotic\\ nature of
the perturbative series.''
A. Vainshtein, 1964 \cite{arkady}}}}

%\end{references}

\end{document}